%% file: arXiv_Culcer_valley_phase_v_5.tex
\documentclass[two column]{revtex4}
\usepackage{graphicx}
\usepackage{color}

\input{abbrev}

\begin{document}

\begin{titlepage}
\centerline{\tt /SET\_team/Neil/other peoples documents/Culcer/valleys 14\_5/manuscript/text/Culcer\_valley\_phase\_v\_5.tex}
\end{titlepage}

\centerline{\today}
\title{Valley Phase and Voltage Control of Coherent Manipulation in Si Quantum Dots}
\author{Neil Zimmerman$^{1}$\footnote{+1-301-975-5887; neilz@mailaps.org; \\  ftp://ftp.nist.gov/pub/physics/neilz/papers.html}, Peihao Huang$^{1,2}$, Dimitrie Culcer\footnote{d.culcer@unsw.edu.au}$^3$}
\affiliation{$^1$Quantum Measurement Division, National Institute of Standards and Technology, MD 20899, USA}
\affiliation{$^2$Joint Quantum Institute, University of Maryland and National Institute of Standards and Technology, MD 20742, USA}
\affiliation{$^3$School of Physics, The University of New South Wales, Sydney 2052, Australia}

\begin{abstract} 

With any roughness at the interface of an indirect-bandgap semiconducting dot, the phase of the
valley-orbit coupling can take on a random value.  This random value, in double quantum dots, causes
a large change in the exchange splitting.  We demonstrate a simple analytical method to calculate
the phase, and thus the exchange splitting and singlet-triplet qubit frequency, for an arbitrary
interface.  We then show that, with lateral control of the position of a quantum dot using a gate
voltage, the valley-orbit phase can be controlled over a wide range, so that variations in the
exchange splitting can be controlled for individual devices.  Finally, we suggest experiments to
measure the valley phase and the concomitant gate voltage control.

\end{abstract}

\maketitle

%
%

%

The physics of valleys in Si is complicated, both theoretically and experimentally.  On the theory
side, calculations of the effects of valleys, especially with rough interfaces, often requires very
large atomistic simulations; those simulations can sometimes make it more difficult to achieve a
simple, intuitive understanding of the underlying physics.  In this paper, we present a very simple
and intuitively-appealing framework to understand and calculate the effect of the phase of the
complex valley-orbit coupling in devices with rough interfaces.  On the experimental side, the
effects of the valleys can be bound up with spin and orbital effects, thus making effective
classical and coherent control of the devices more challenging.  We propose a simple experimental
method (lateral gate voltages) to both analyze and to control these confounding experimental effects
when they arise from the complex phase.

Quantum coherent manipulation of electrons in Si has been a very active field of study
recently\cite{Zwanenburg13a}\cite{Morton11a}.  The advantages of Si include low spin-orbit coupling
and the ability to isotopically enrich $^{28}$Si, both of which will tend to reduce the decoherence
of spin qubits.  In addition, the integration with CMOS classical circuitry holds the potential for
monolithically integrating qubits with control circuits\cite{Morton11a}.  Very recent advances in
demonstrating quantum coherence include driving single spins in SiGe quantum dots with
micromagnets\cite{Scarlino15a}\cite{Kawakami14a}, single spins in Si/SiO$_2$ quantum dots including
Stark-shifted addressability\cite{Veldhorst14a}, and a two-qubit gate in Si/SiO$_2$ quantum
dots\cite{Veldhorst15a}.

One complication for quantum control in Si is engendered by the valley degree of freedom; there are
six equivalent positions in the three-dimensional band structure\cite{Zwanenburg13a}.  Because there
is no external control of the valley quantum number, complications include difficulty in
well-defined initialization and enhanced spin decoherence at valley relaxation
``hotspots"\cite{Yang13a}.  Experimental measurements of the magnitude of the valley splitting
between the lowest two valley states include spin-valley "hotspots"\cite{Hao13a} and valley
splitting magnitude tunable with electric field\cite{Yang13a}\cite{Goswami07a}.  The experimental
work most relevant for our study is Shi et al\cite{Shi11a}, which showed experimentally a shift in
the single-triplet splitting $J$ for two electrons {\it on a single dot}, where $J$ is dominated by
the energy difference between single-particle ground and excited valley-orbit states in Si/SiGe.  In
particular, they measured $J$ as a function of lateral shift using gate voltages; they showed about
a 20 \% shift, and interpreted it as coming from the change in single-particle valley-orbit energies
arising from a rough interface (they solved this for the toy model of a single atomic step).

In the vicinity of an interface, the lowest energy valleys are typically perpendicular to the
interface, and are split by the valley-orbit coupling, which is defined as
\begin{equation}
{\Delta} = \tbkt{z}{(\mathcal{V} + eFz)}{\bar{z}},
\end{equation}
where\ket{z}, \ket{\bar{z}} are the bare valley states centered at $k_z = \pm k_0 = \pm (0.85)
2\pi/a_0$, $\mathcal{V}$ is the interfacial potential energy, $F$ is the applied electric field, and
$a_0$ is lattice constant; this leads to eigenstates which are symmetric superpositions of $\ket{z}$
and $\ket{\bar{z}}$ (see text above Eqn \ref{tsubminus}), split by $|\Delta|$.  We note that, by
virtue of the definition, the coupling is a complex quantity $\Delta = |\Delta| e^{i \phi}$.  

Theoretical predictions of the valley-orbit coupling include predictions of magnitude in perfect
SiGe\cite{Boykin04a}\cite{Boykin04b}\cite{Boykin06a}\cite{Friesen10a} and
SiO$_2$\cite{Saraiva09a}\cite{Saraiva11a}, magnitude in disordered
interfaces\cite{Srinivasan08a}\cite{Gamble13a}\cite{Culcer10b}, effects of valley phase on
magnitude\cite{Friesen07a}\cite{Friesen06a}\cite{Wu12a}, spin-valley interactions\cite{Culcer10a}\cite{Huang16a} and intervalley dynamics at interface steps\cite{Boross16a}.

In contrast to the magnitude, there has been somewhat less consideration of the complex
phase\cite{Friesen07a}\cite{Friesen06a}\cite{Wu12a}\cite{Saraiva09a} $\phi$.  In terms of the effect
of interface roughness on tunneling, Shiau et al \cite{Shiau07a} used an atomistic tight-binding
model to calculate, among other quantities, the intra- and intervalley tunneling matrix elements,
between lead and dot, as a function of atomic step position and miscut angle; they also briefly
discussed the significance of the valley phase in preserving valley index under tunneling.
Similarly, Gamble et al \cite{Gamble13a} used a semianalytical expansion in disorder matrix elements
to calculate, among other quantities, the intervalley matrix element, between two dots, as a
function of atomic step position.

Although the global phase $\phi$ for a single quantum dot or donor has no physical consequence, the
phase difference $\Delta\phi$ between two quantum dots or donors has great significance, because
matrix elements (such as exchange splitting\cite{Wu12a} and preservation of valley index
\cite{Shiau07a}) between the two dots depend on $\Delta\phi$.  As in studies of the magnitude of
the valley splitting, the roughness of the interface is the crucial device feature; this is
because the Bloch wavefunction is tied to the Si lattice, while the envelope of the electron
wavefunction follows the rough interface.  With $d = a_0/4$ the thickness of a single terrace
step, $2 k_0 d \approx 2 (0.85) (2\pi/a_0) (a_0/4) \approx 0.42 (2\pi)$ [ref\cite{Friesen07a}].  Thus, not only is the
value of $\phi$ large, it is also incommensurate with $2\pi$; this basic fact is the origin of the
large effects discussed in this paper.

The main point of this paper is to present a simple analytic approach to estimating the effects of
interface roughness on qubit gate operation frequencies through $\Delta \phi$ and $J$.  We do this
by focussing on the valley phase as an intermediary to calculate the exchange energy for two
electrons in two quantum dots.  This occurs because, as we will show, the exchange splitting $J
\propto 1 + \cos \Delta\phi$.  Thus, the large value of $\Delta\phi$ manifests itself as a large
effect on the $\sigma_z$\ rotation frequency ($J/h$) for both two electrons on two dots (2e2d)
singlet-triplet qubits and the two-qubit gate coupling frequency ($J/h$) for 2e2d single spin
qubits; as we will show, these effects are present for both interfaces with steps (corresponding to
Si/SiGe) and with random fluctuations (Si/SiO$_2$).

In this paper, we will show (within the effective mass approximation) i) the derivations of
$\Delta\phi$ based on interface roughness and the dependence of $J$ on $\Delta\phi$, ii) results of
simulations of $\Delta\phi$ for both step and random interfaces, and iii) the ability to control the
coupling phase with applied gate voltage, by laterally shifting quantum dots.  {\it We emphasize
  that, in contrast to previous results}\cite{Shiau07a}\cite{Gamble13a} {\it where tunneling matrix
  elements have been obtained for some particular interface roughnesses, the power of the present
  work is combining the focus on the valley phase with the resulting very simple analytical method
  to calculate the effect on tunneling and valley index preservation of any arbitrary interface
  roughness.}  For comparison, we note that a full atomistic calculation for a single dot typically
requires 1 million atoms and at least 10 orbitals per atom\cite{Rahman2012a}; in addition, for
multiple dots, the number of atoms would increase substantially to take into account both the
additional dots and the large volume between the dots.

We consider two electrons, one in each of two quantum dots (L and R denote the two lateral locations
of the quantum dots), localized in the z-direction by the combination of interface and accumulation
gate voltage, and localized in the x-, y-directions by lateral gates.  For each dot, the potential
is then
\begin{equation}\label{potential}
V_L (x, y, z) = \frac{\hbar^2}{2m^*a^2} \, \bigg[ \frac{(x - x_L)^2 + y^2}{a^2} \bigg] + \mathcal{V} (x,y,z) + eFz,
\end{equation}
and similarly for position R.  The QD has a Fock-Darwin radius $a$, with $m^*$ the Si in-plane
effective mass.  The effective mass (EMA) single-electron wavefunction for a multi-valley system
(left dot L and valley z) is
\begin{equation}\label{wavefunction}
L_z (x, y, z) = \phi_L(x,y) \, \psi (z) \, u_z ({\bf r})\, e^{ik_0 z},
\end{equation} 
and similarly for position R and valley $\bar{z}$; here, $\psi(z)$ is
the envelope function for the $z$-direction and $u_z ({\bf r})$ has
the periodicity of the lattice.  Then, in the absence of valley-orbit coupling, the lowest orbital
state is
\begin{equation}
\begin{array}{rl}
\phi_L (x,y) &=\frac{1}{a\sqrt{\pi}} \, e^{-\frac{(x - x_L)^2+y^2}{2a^2}},
\end{array}
\end{equation}
and similarly for position R.

Finally, the interface potential is 
\begin{equation}\label{potential}
\begin{array}{rl}
\displaystyle \mathcal{V} (x,y,z) = & \displaystyle U_0 \, \theta[\zeta(x,y) - z], 
\end{array}
\end{equation}
where $\zeta(x, y)$ is the height of the rough interface.  With this interface, we make the
assumption that the center-of-charge of the wavefunction follows exactly: $\psi (z) \rightarrow \psi(z -
\zeta)$.  This assumption is justified when the variation of $\zeta(x, y)$ with respect to $x, y$ is
not too large (see Supplement).  Combining this with Equations 1, 3, \ref{potential}, we obtain
\begin{equation}\label{DeltaRough}
\Delta_D \approx \Delta_0 \int_{-\infty}^{\infty} \int_{-\infty}^{\infty} dx\, dy\, |\phi_D(x, y)|^2
\, e^{-2i k_0 \zeta(x, y)},
\end{equation}
where $\Delta_0$ is the valley splitting in the absence of interface roughness.  Equation
\ref{DeltaRough} is simple and appealing: It shows a simple analytical connection between the
interface roughness $\zeta(x, y)$ and the valley phase\cite{Friesen07a}.  This simple result depends
crucially on only two assumptions: i) the electron wavefunction locally follows the rough interface
(Supplement A. through C., including particularly an analysis based on comparing kinetic energy cost
to potential energy gain) ); ii) the electron wavefunction contains approximately only the ground
orbital state (Supplement D).

Finally we define
$\Delta_{L, R} = |\Delta_{L, R}| e^{-i \phi_{L, R}}$ and $\Delta\phi = \phi_L - \phi_R$.  We note
that the assumption of pure position states L and R is equivalent to assuming the Coulomb repulsion
energy is large compared to the tunneling energy, which is generally experimentally valid.  A simple
consequence of Equation \ref{DeltaRough}: If we consider a flat interface with one step of height
$d$ located between the two dots, because $\zeta$ is defined with respect to the underlying Si
lattice, $\Delta\phi = 2 k_0 d$, as discussed above.

We now turn to the dependence of exchange energy on valley phase: After mixing $\ket{z}$ and
$\ket{\bar{z}}$ through the valley splitting, the single-particle states are $\ket{L_\pm} =
\frac{1}{\sqrt{2}} (\ket{L_z} \pm e^{i\phi_L}\ket{L_{\bar{z}}}$, and similarly for R.  Thus, the
tunneling matrix element between lowest-lying states $t_{--} = t_{++} = \tbkt{L_{-}} {T} {R_{-}}$ is
\begin{equation}\label{tsubminus}
\begin{array}{rl}
t_{--} = & \frac{1}{2} (\bra{L_z} - e^{-i \phi_L} \bra{L_{\bar{z}}}) T (\ket{R_z} - e^{i \phi_R} \ket{R_{\bar{z}}}) \\
[3ex]
= & \frac{1}{2} [ \tbkt{L_z}{T}{R_z} + e^{-i \Delta\phi} \tbkt{L_{\bar{z}}}{T}{R_{\bar{z}}}] = \frac{t_0}{2} [1 + e^{-i \Delta\phi}],
\end{array}
\end{equation}
where $T$ and $t_0$ are the tunneling Hamiltonian and the tunneling matrix element in the absence of
valley effects, respectively.  Similarly, $t_{-+} = \frac{t_0}{2} [1 - e^{-i \Delta\phi}]$.
  The exchange energy $J = 4 |t_{--}|^2/U$, where $U$ is the on-site repulsion energy, is thus
\begin{equation}\label{Jfinal}
J = \frac{J_0}{2} (1 + \cos \Delta\phi),
\end{equation}
where $J_0 = 4 |t_{0}|^2/U$ is the exchange energy in the absence of valley phase.

\begin{figure}[htbp] 
\includegraphics[clip=true, viewport=0.5in 1.2in 5in 7.0in,scale=0.8]{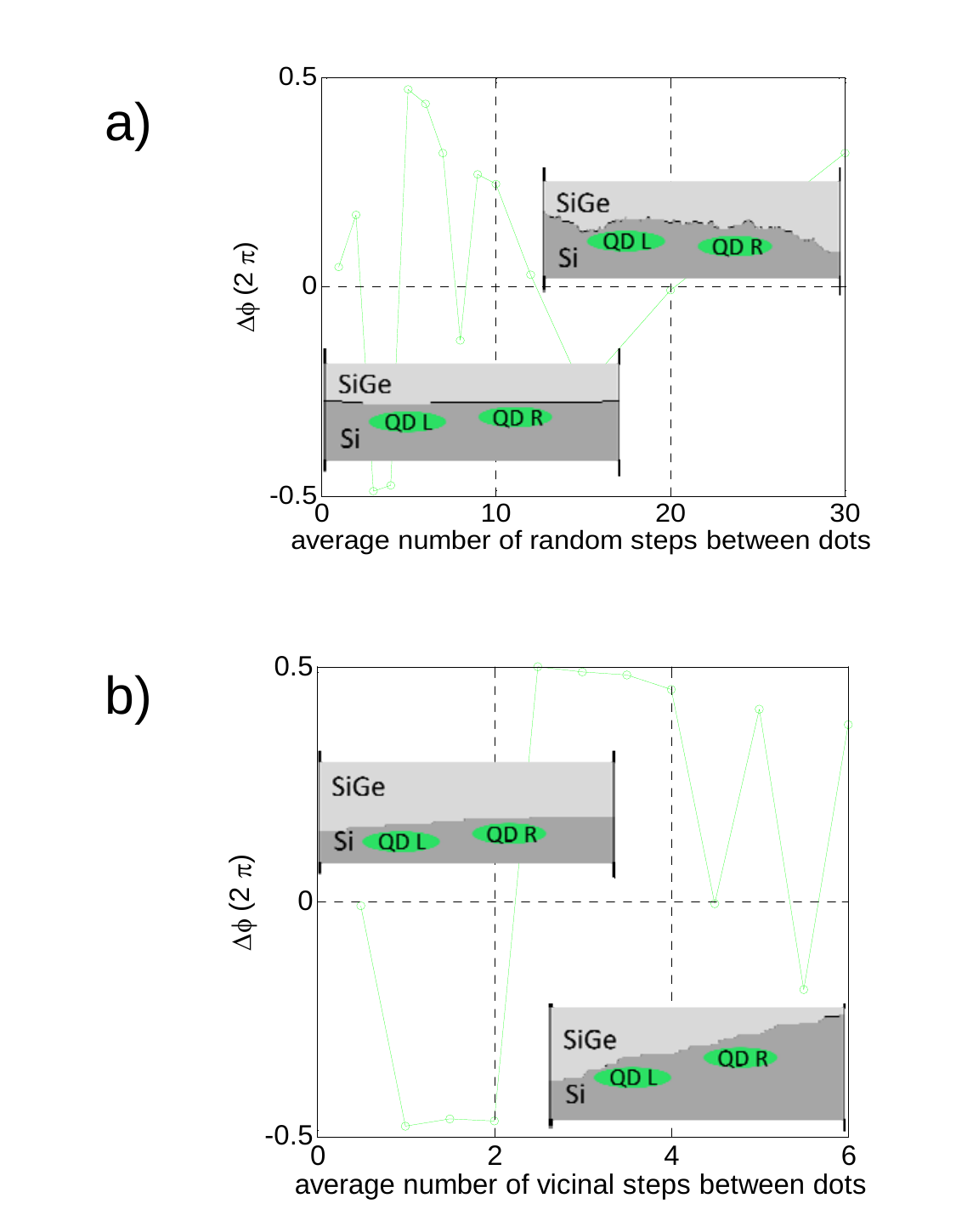}
\caption{\label{fig:SiGe} Calculation of $\Delta\phi $ for Si/SiGe interfaces (horizontal sections
  separated by vertical steps), for two types of simulated interfaces: a) Interface is on average
  horizontal, and contains upgoing and downgoing steps randomly distributed with a lateral density
  as indicated; b) Interface is tilted (vicinal), with only upgoing steps randomly distributed as
  indicated.  Insets are sketches of representative interfaces at the two ends of the ranges of
  lateral densities.  Note that $\Delta\phi$ is large for all rough interfaces, because $2 k_0\ a_0/4$
  is large.}
\end{figure}

Figure \ref{fig:SiGe} shows the results for simulated SiGe interfaces (horizontal sections separated
by vertical steps) for both random (approximately equal numbers of upgoing and downgoing steps) and
for vicinal (all steps are upgoing) interfaces.  The interfaces were generated by assigning steps at
random lateral positions so as to get a particular average lateral density, as indicated in the
graphs.  As shown, for any roughness, $\Delta\phi $ is of large magnitude compared to $2\pi$, and
has no systematic dependence on tilt or randomness; these facts both occur because, as discussed
above, $2k_0d$ is large for a single terrace step.

\begin{figure}[htbp] 
\includegraphics[clip=true, viewport=0.5in 1.2in 5in 7.0in,scale=0.8]{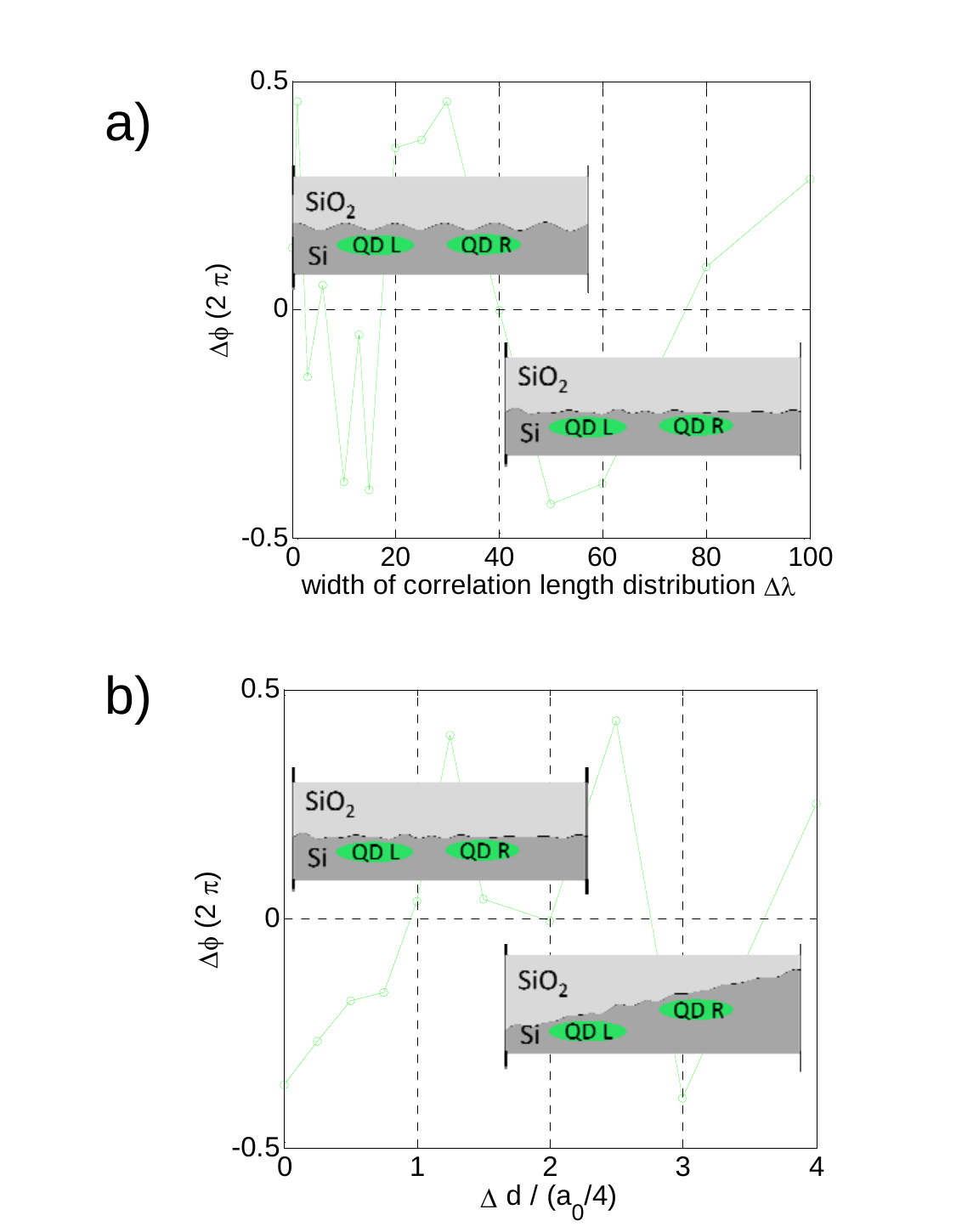}
\caption{\label{fig:SiO2} Calculation of $\Delta\phi$ for Si/SiO$_2$ interfaces: a) Interface is on
  average horizontal, and is generated by inverse Fourier transform of a randomly generated function
  of variable width.  Larger width yields a more random interface, as indicated by the inset
  sketches.  b) Similar to a), using a single random interface and a variable amount of vicinal
  tilt; $\Delta d$ is the vicinal height difference between the two dots, so that $\Delta d /
  (a_0/4)$ is the height difference in units of equivalent terrace steps (compare to Figure
  \ref{fig:SiGe}).  Note that the interface is not flat at the left end of both graphs, so that
  $\Delta\phi$ is nonzero there.}
\end{figure}

Similar to Figure \ref{fig:SiGe}, Figure \ref{fig:SiO2} shows the calculation for $\Delta\phi$
appropriate for a Si/SiO$_2$ interface without and with vicinal tilt.  We generated this interface
by i) first generating a random Fourier transform of variable width $\Delta\lambda$, ii) inverse
Fourier transforming, and then iii) scaling to give a total vertical height of about 1.5 nm
(equivalent to the height of 10 random steps in Figure \ref{fig:SiGe}a).  In Figure
\ref{fig:SiO2}a), zero width corresponds to a single correlation length of 20 nm (sinusoidal
roughness), and the largest width $\Delta\lambda = 100$ corresponds to a range of correlation
lengths between two and 20 nm.  Previous high resolution TEM and weak localization studies of the
Si/SiO$_2$ interface have yielded a range of correlation lengths from a few nm\cite{ikarashi00a} up
to 100 nm\cite{krivanek80a}\cite{anderson93a}.  We see that, as for the SiGe interface, vicinal and
non-vicinal randomness with a variety of correlation lengths all lead to large values of
$\Delta\phi$.

We now turn to considering an experimental control of $\Delta \phi$ and $J$, using gate voltages.
Since the valley phase in Equation \ref{DeltaRough} arises from the vertical height of the
interface, at first glance one might think that moving the dot vertically into the Si substrate
might have a strong effect on $\Delta\phi$; although a vertical electric field has strong effect on
the magnitude of the coupling$^{[9] - [21]}$, it has a
small effect on $\Delta\phi$\cite{Saraiva09a}, at least for a flat interface.  However, due to the strong dependence
of the valley phase on the exact local realization of the interface roughness, using a gate voltage
to shift the dot laterally can have a large effect.

\begin{figure}[htbp] 
\includegraphics[clip=true, viewport=0.5in 1.4in 7in 6.7in,scale=0.66]{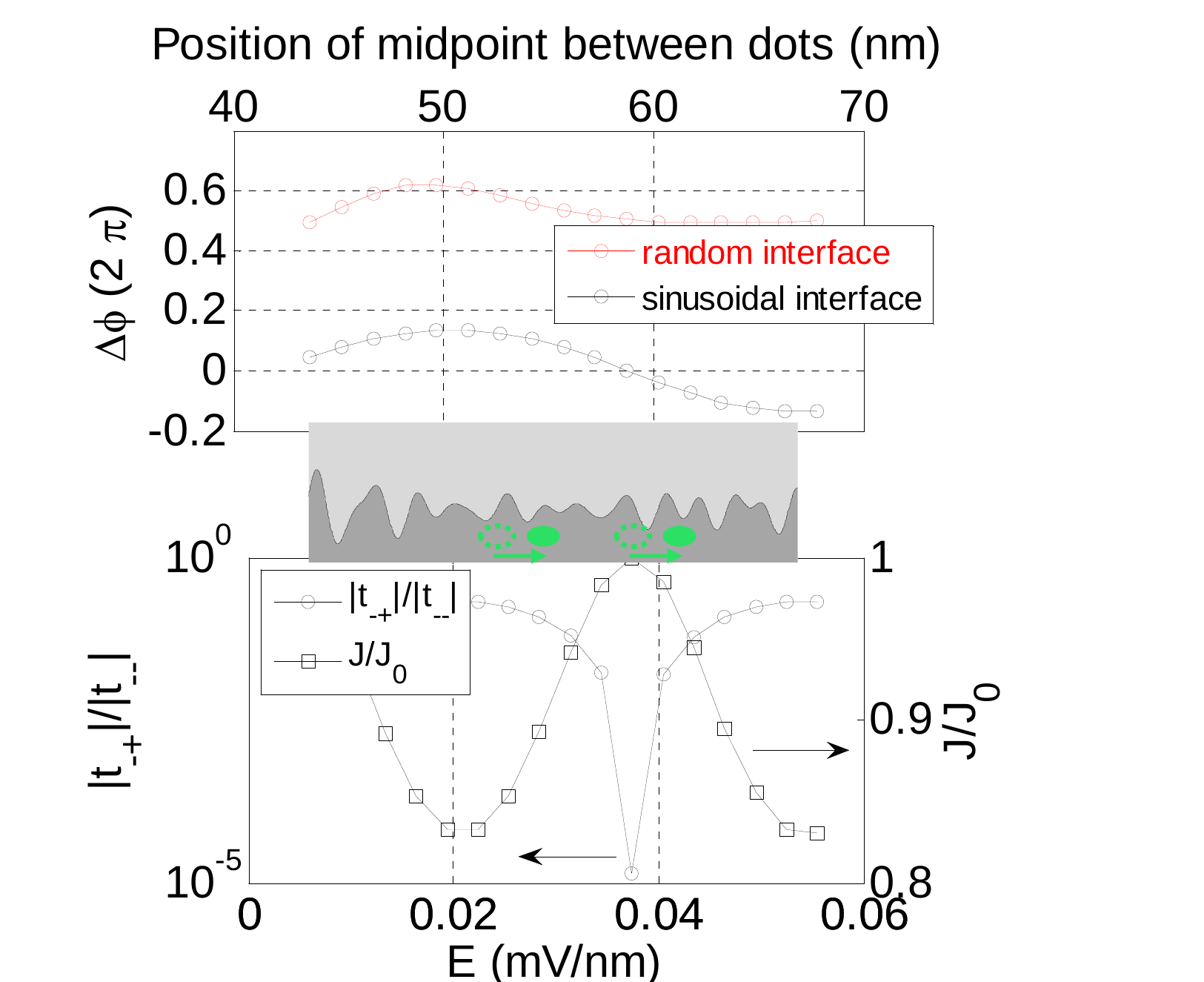}
\caption{\label{fig:voltage} Using lateral gate voltages to control $\Delta \phi$ and $J$ by moving
one or both quantum dots.  Lower panel: $J$ and $t_{-+}$ are from sinusoidal interface result in upper
plot.  In this simulation, the separation between the dots is held fixed, to maintain the bare
tunneling energy $t_0$ as a constant.  Lateral electric field values are estimated for a parabolic
confining potential that would yield a quantum dot of orbital size 60 nm and energy spacing 1 meV}
\end{figure}

Overall, Figure \ref{fig:voltage} shows this possibility: the upper panel shows the effect on
$\Delta\phi$ of using a lateral gate voltage to move both dots laterally with a constant separation,
for both a sinusoidal interface ($\Delta\lambda = 0$) and random interface ($\Delta\lambda = 10$).
Clearly, the phase difference $\Delta\phi$ and the concomitant energies are strong functions of the
lateral position; this is simply because moving a dot changes the local interface height, and thus
the phase factor in Equation \ref{DeltaRough}.  The lower panel shows the concomitant effect on the
tunneling matrix elements and on the exchange energy.  While the specific gate voltage dependence is
likely impossible to predict in advance, this Figure shows that there is the hope of operationally
controlling qubit gate frequencies; this experimental control
does not depend on our simple analytical framework\cite{Shiau07a}\cite{Gamble13a}.  In comparing
Figure \ref{fig:voltage} to Figures \ref{fig:SiGe} and \ref{fig:SiO2}, we note that the systematic
appearance of Figure \ref{fig:voltage} relies on the fact that the interface in that Figure has a
correlation length approximately equal to the total range of 20 nm, so that the random structure in
$ \Delta\phi$ apparent in Figures \ref{fig:SiGe} and \ref{fig:SiO2} only appears over the total
range of separation in Figure \ref{fig:voltage}.

How can we test the complex valley phase experimentally?  First of all, it appears that the global
phase (i.e., $\phi_L$) is not measurable, because there is no physical realizable corresponding to
it.  We suggest two possible experiments for testing $\Delta\phi$ directly: i) Measurements of
the exchange energy splitting $J$ (references \cite{Weber14a}\cite{Petta05a}) in a DQD could be done
as a function of multiple lateral gate voltages in the same way as described above.  If $J$ changes
in a systematic way with the lateral position of both dots, this would indicate the likelihood of
valley phase-mediated modulation of $J$; ii) Very recently, valley blockade in a double quantum dot
(DQD), exactly analogous to Pauli spin blockade\cite{Lai11a}\cite{Shaji08a}\cite{Weber14a}), has
been reported\cite{Perron16a}.  In this work, an asymmetry in the size of the bias triangles in a
DQD is attributed to blockade in one bias direction due to the impossibility of inter-valley
tunneling.  Such valley blockade depends on $\Delta\phi = 0$, and thus the leakage current in the
blockaded region $\sub{I}{leak} \propto (1 - \cos \Delta\phi) / 2$.  Thus, we suggest examining the
leakage current magnitude as a function of lateral gate voltage.

Also, we estimate as insignificant the magnitude of spin dephasing (through the dependence of $J$ on
$\Delta\phi$) due to charge noise which would modulate the lateral position of one or both dots:
From Figure \ref{fig:voltage}, $dJ/dx \approx 0.1 J_0 / $(10 nm), lateral motion $dx/dV_G \approx$
(1 nm/10 mV), and a voltage noise magnitude $\Delta V_G = 1\ \mu$V, we obtain the noise magnitude
$\Delta J \approx 10^{-6} J_0$.  Finally, one way of suppressing the complex valley phase effects
and of increasing the valley splitting magnitude, would be to make devices with perfectly flat
atomically sharp terraces as large as $10 \mu$m\cite{Li11a}\cite{Tanaka96a}.  However, it is likely
that subsequent growth of SiO$_2$ or SiGe will lead to substantial roughening of such terraces, and
thus it appears likely that these valley phase effects will still be present in typical quantum
coherent devices.

It is a pleasure to acknowledge useful discussions with Josh Pomeroy, Mark Stiles, Michael Stewart Jr
(all of NIST), Andras Palyi (BUTE), John Gamble (Sandia) and Chris Richardson (LPS).

%
%
\bibliography{ref_QI_08}

\clearpage

{\bf Supplementary Information on Whether the Electron Wavefunction Follows the Interface}

In this Section, we will discuss this question as follows:

\begin{enumerate}

	\item We will give a formal analysis, based on symmetry arguments, that yields the result
          that the wavefunction follows the interface for any interface slope less than one.  This
          analysis is relevant for the smooth (Si/SiO$_2$) interfaces considered in this paper.
	\item We will give an estimate based on energy considerations, that yields the result that
          the wavefunction follows the interface for any interface roughness with correlation length
          greater than about 0.3 nm.  This analysis is relevant for both step (Si/SiGe) and smooth
          (Si/SiO$_2$) interfaces considered in this paper.
	\item Based on the two analyses, we will present a result from our simulations where we
          compare the valley phase for i) an interface with vertical steps to ii) an interface with
          sloped steps, and show that the valley phase is approximately the same for both
          interfaces.  This result shows that the formal analysis (item 1. above) is thus also
          relevant for the step (Si/SiGe interfaces).

\end{enumerate}

\subsection{Symmetry Arguments}

The Schrodinger equation in the absence of interface roughness is
\begin{equation}
\bigg[-\frac{\hbar^2\nabla^2}{2m} + V ( {\bf r}) \bigg] \, \sub{\psi}{ew} ({\bf r}) = \varepsilon \, \sub{\psi}{e} ({\bf r}),
\end{equation}
where \sub{\psi}{e} is the total electron wavefunction (represented as $L_z (x, y, z)$)
in Equation \ref{wavefunction} in the main text.  The solution to this equation is
$\sub{\psi}{e, 0}({\bf r})$, the solution for a flat interface.

We wish to examine the effect on \sub{\psi}{e} when the interface becomes rough, and in
particular on the amount by which the spatial derivatives change.  We can define new coordinates
\begin{equation}
\begin{array}{rl}
\displaystyle x' = & \displaystyle x \\ [3ex]
\displaystyle y' = & \displaystyle y \\ [3ex]
\displaystyle z' = & \displaystyle z - \zeta(x, y).
\end{array}
\end{equation}
The differentials transform as
\begin{equation}
\begin{array}{rl}
\displaystyle \pd{\sub{\psi}{e}}{x} = & \displaystyle \pd{\sub{\psi}{e}}{x'}\pd{x'}{x} + \pd{\sub{\psi}{e}}{z'}\pd{z'}{x} = \pd{\sub{\psi}{e}}{x'} + \pd{\sub{\psi}{e}}{z'}\pd{z'}{x} \\ [3ex]
\displaystyle \pd{\sub{\psi}{e}}{y} = & \displaystyle \pd{\sub{\psi}{e}}{y'}\pd{y'}{y} + \pd{\sub{\psi}{e}}{z'}\pd{z'}{y} = \pd{\sub{\psi}{e}}{y'} + \pd{\sub{\psi}{e}}{z'}\pd{z'}{y} \\ [3ex]
\displaystyle \pd{\sub{\psi}{e}}{z} = & \displaystyle \pd{\sub{\psi}{e}}{z'}\pd{z'}{z} = \pd{\sub{\psi}{e}}{z'}.
\end{array}
\end{equation}
The second differentials become
\begin{widetext}
\begin{equation}
\begin{array}{rl}
\displaystyle \pd{}{x}\pd{\sub{\psi}{e}}{x} = & \displaystyle \bigg(\pd{}{x'} + \pd{z'}{x}\pd{}{z'}\bigg) \bigg( \pd{\sub{\psi}{e}}{x'} + \pd{\sub{\psi}{e}}{z'}\pd{z'}{x} \bigg) = \bigg[ \frac{\partial^2\sub{\psi}{e}}{\partial x^{'2}} + 2\frac{\partial^2\sub{\psi}{e}}{\partial x' \partial z'}\pd{z'}{x} + \frac{\partial^2\sub{\psi}{e}}{\partial z^{'2}} \bigg(\pd{z'}{x}\bigg)^2 \bigg]  \\ [3ex]
\displaystyle \pd{}{y}\pd{\sub{\psi}{e}}{y} = & \displaystyle \bigg(\pd{}{y'} + \pd{z'}{y}\pd{}{z'}\bigg) \bigg(\pd{\sub{\psi}{e}}{y'} + \pd{\sub{\psi}{e}}{z'}\pd{z'}{y}\bigg) = \bigg[ \frac{\partial^2\sub{\psi}{e}}{\partial y^{'2}} + 2\frac{\partial^2\sub{\psi}{e}}{\partial y' \partial z'}\pd{z'}{x} + \frac{\partial^2\sub{\psi}{e}}{\partial z^{'2}} \bigg(\pd{z'}{y}\bigg)^2 \bigg] \\ [3ex]
\displaystyle \pd{}{z}\pd{\sub{\psi}{e}}{z} = & \displaystyle \frac{\partial^2\sub{\psi}{e}}{\partial z^{'2}}.
\end{array}
\end{equation}
\end{widetext}
Thus, the second derivatives in the original coordinate system are approximately the same
as those in the new system, if $\displaystyle \bigg|\pd{z'}{x, y}\bigg| \ll 1$.  In this
case, we can immediately see that $\sub{\psi}{e} (x, y, z) \approx \sub{\psi}{e, 0} (x, y,
z - \zeta (x, y))$.  In summary, if the slope of any interface roughness is substantially
less than one, the electron wavefunction will approximately follow the interface.

\subsection{Energy Considerations}

Here, we consider the interplay between kinetic and potential energies.  For a rough interface,
since the accumulation gate produces a potential well in the z-direction at the interface and
follows conformally the interface roughness, the electron potential energy will be lowered if the
wavefunction follows the interface roughness exactly (Fig. S2).  On the other hand, the more the
wavefunction follows the interface roughness, the larger is the kinetic energy increase.  For
simplicity, we consider only one lateral direction $x$.

Thus, we determine the approximate maximum bending of the electron wavefunction by the constraint

\begin{equation}\label{energies}
 \frac{\hbar^2}{2m} \frac{d^2\sub{\psi}{ew}}{d x^2} < (\mathcal{V} + eFz)\sub{\psi}{ew}
\end{equation}

in the effective mass approximation (EMA).  Note that, for this inequality, the normalization of the wavefunction is present on both sides, so we will suppress the normalization factor.

\begin{figure}[htbp] 
\includegraphics[clip=true, viewport=0.5in 1.2in 7in 6.0in,scale=0.5]{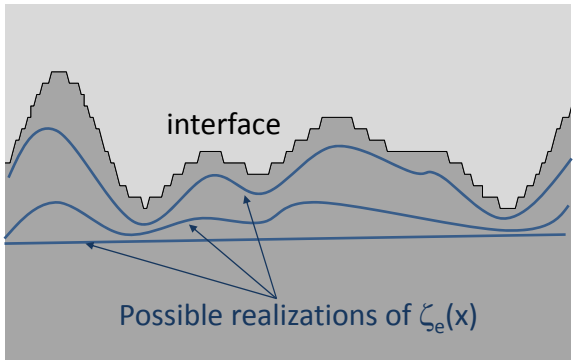}
\caption{\label{fig:zeta_e} Figure S1: Possible realizations of \sub{\zeta}{ew} (x); the correct
  realization depends on the interplay between kinetic and potential energy.  Bottom curve: kinetic
  energy term dominates; top curve: potential energy term dominates.
}
\end{figure}

We define the following parameters:

\begin{description}

	\item[$\sub{\psi}{ew} = \phi(x,y) \, \psi (z)$] Electron envelope wavefunction in the EMA.
	\item[\sub{\zeta}{ew} (x)] Height of center of electron wavefunction (as distinguished from
          the height of the interface $\zeta (x)$).  [See Figure \ref{fig:zeta_e}]
	\item[\sub{\zeta}{e0}] Amplitude of local bending of electron wavefunction (assumed to be quadratic).  
	\item[\sub{l}{ce}] Correlation length of electron wavefunction (as distinguished from the correlation length of the interface).
	\item[\sub{U}{0}] Band offset between Si and SiO$_2$, as in Equation \ref{potential} of the main text.
	\item[$t$] Vertical thickness of electron 2DEG (inversion layer thickness).
	\item[$m$] Effective mass of transverse electron in Si; $m = 0.2  \sub{m}{e}$.

\end{description}

\subsubsection{Kinetic Energy}

To approximate the left-hand side of Equation \ref{energies}, we start by assuming that the wavefunction is locally quadratic:

\begin{equation}
\sub{\zeta}{ew} (x) = (\sub{\zeta}{e0}) \large ( \frac{x}{\sub{l}{ce}} \large)^2.
\end{equation}

Following Equation \ref{wavefunction} in the main text, the EMA envelope wavefunction is

\begin{equation}
\sub{\psi}{ew} (x, z) = \phi_D(x) \, \psi (\sub{\zeta}{ew} (x) - z).
\end{equation}

A good approximation to the z-dependence of the 2DEG is $\psi(x, z) = e^{-(\sub{\zeta}{ew}
  (x) - z)/t}$ (ref \cite{Streetman95a}).  The contribution to the kinetic energy from the
roughness of the 2DEG comes only from this part of the total wavefunction \sub{\psi}{ew} (x, z):

\begin{equation}
\frac{d^2\sub{\psi}{ew}}{dx^2} = \sub{\psi}{ew} [\large (\frac{\sub{\zeta}{ew}^{\prime}}{t}\large )^2 - \frac{\sub{\zeta}{ew}^{\prime\prime}}{t}].
\end{equation}

With the quadratic dependence of \sub{\zeta}{ew} (x), and evaluating at $x =
\sub{l}{ce}$, we obtain

\begin{equation}
\frac{d^2\sub{\psi}{ew}}{dx^2} \approx \sub{\psi}{ew}  \large (2 \frac{\sub{\zeta}{e0}}{\sub{l}{ce} t}\large )^2[1 -
  \frac{t}{2 \sub{\zeta}{e0}}].
\end{equation}

With $t/\sub{\zeta}{e0}  \approx$ 5 nm / 0.5 nm, 

\begin{equation}
\frac{d^2\sub{\psi}{ew}}{dx^2} \approx \sub{\psi}{ew}  (-2 \frac{\sub{\zeta}{e0}}{\sub{l}{ce}^2 t}).
\end{equation}

Finally, the left hand side of Equation \ref{energies} is

\begin{equation} \label{kinetic_energy}
\frac{\hbar^2}{2m} \frac{d^2\sub{\psi}{ew}}{d x^2}  \approx \sub{\psi}{ew}  (-\frac{\hbar^2}{m} \frac{\sub{\zeta}{e0}}{\sub{l}{ce}^2 t}).
\end{equation}

\subsubsection{Potential Energy}

We approximate the vertical potential well at the interface as triangular, with the sloped section having a slope of $\sub{U}{0} / t$.  The decrease in potential energy when the electron wavefunction is bent with a vertical change of approximately \sub{\zeta}{e0} instead of flat is thus

\begin{equation}\label{potential_energy}
(\mathcal{V} + eFz)\sub{\psi}{ew}  \approx \sub{\psi}{ew}   \frac{\sub{U}{0} \sub{\zeta}{e0}}{t}.
\end{equation}

\subsubsection{Summary}

Combining Equations \ref{energies}, \ref{kinetic_energy}, \ref{potential_energy}, we thus obtain a constraint on the minimum radius of curvature or correlation length of the electron wavefunction

\begin{equation}
\sub{l}{ce}^2 > \frac{\hbar^2}{m \sub{U}{0}};
\end{equation}

using the values in this section ($U_0 =$ 3 eV for Si/SiO$_2$), we finally obtain $\sub{l}{ce}
>\approx$ 0.3 nm.  Given that the typical amplitude of roughness is less than or of order 1 nm, this
condition is quite similar to that in the previous argument based on symmetry considerations.

\subsection{Results from Simulations}

\begin{figure}[htbp] 
\includegraphics[clip=true, viewport=0.5in 1.2in 7in 6.0in]{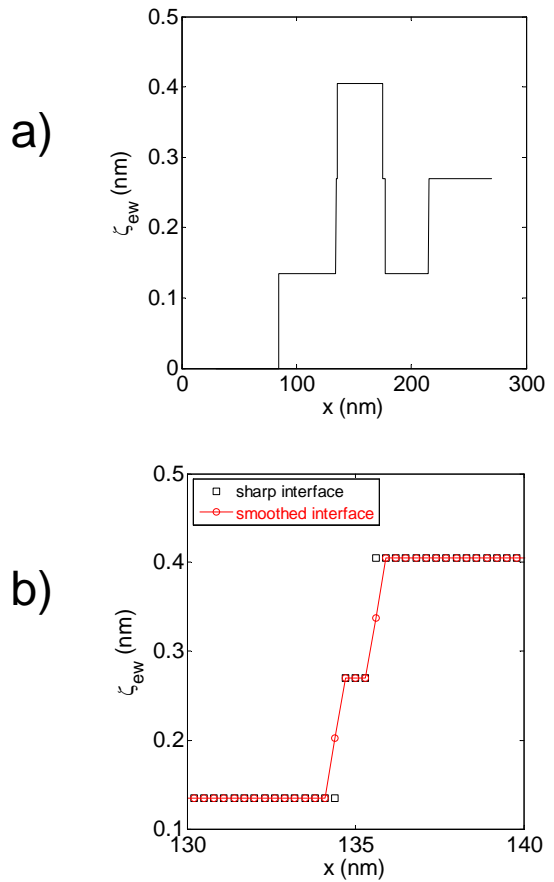}
\caption{\label{fig:smoothed} Figure S2: Sharp and smoothed vertical centers
  \sub{\zeta}{ew} (x) of electron wavefunction, for testing dependence of $\Delta\phi$ on
  the extent to which electron wavefunction follows interface roughness.  a) Full extent
  of sharp \sub{\zeta}{ew} (x) for Si/SiGe simulated interface; quantum dots are centered
  at 125 and 175 nm, with dot radius $a = 10.4$ nm;  b) Small
  region of interface, showing smoothing of the original sharp \sub{\zeta}{ew} (x) to give
  a slope less than one, or a correlation length greater than 0.3 nm. 
}
\end{figure}

Figure  \ref{fig:smoothed} shows two choices for \sub{\zeta}{ew} (x), with the smoothed one obeying the approximate constraints derived in the last two sections.  We simulated the phase difference, and got the result

\begin{equation}
\Delta\phi_{sharp} = -0.110 (2\pi), \Delta\phi_{smoothed} = -0.111 (2\pi).
\end{equation}

Thus, the results of a combination of i) two theoretical constraints on the sharpness of the electron wavefunction as viewed through \sub{\zeta}{ew} (x), and ii) the very small phase difference for the sharp versus smoothed electron wavefunctions, demonstrate that our assumption in the main text (that the electron wavefunction follows the interface roughness) does not present a significant offset in our results.

\subsection{Using only the Ground Orbital State}

In this section, we give the assumption underlying the use of only the ground orbital state, and discuss the plausibility of this assumption.

For a flat interface, we can identify the eigenstates as $\ket{\lambda z} = \phi_\lambda 
  (x,y) \, \psi (z) \, u_z ({\bf r})\, e^{ik_0 z}$, following Equation \ref{wavefunction} in the
  main text.  Here, $\lambda$ is the orbital state index, and $z$, $\bar{z}$ are the two possible
  valley states.

For the full treatment of the rough interface, we would consider a potential $V = \sub{V}{flat} +
\Delta V$, where \sub{V}{flat}\ is the potential in Equation \ref{potential} in the main text, and
$\Delta V$ is a perturbation on the flat interface potential.  In this case, in order to calculate
the full eigenstates, we would consider Hamiltonian matrix elements $\tbkt{\lambda z}{\Delta
  V}{\lambda' \bar{z}}$, including intra-valley intra-orbital, intervalley intra-orbital, and
intervalley inter-orbital terms.  The first two types of terms, while leading to mixing of the
orbital states, will not affect our results for the valley phase.

However, the last term (intervalley and inter-orbital) will affect, and in particular will likely
make substantially smaller through statistical averaging, the valley phase.  The relevant parameter
that determines the level of intervalley interorbital mixing is $\beta = \frac{\tbkt{\lambda
    z}{\Delta V}{\lambda' \bar{z}}}{E_\lambda - E_\lambda'}$.

Thus, the assumption underlying the use of only the ground orbital states is that $\beta << 1$.  There are two arguments for believing that this approximation is justified:

\begin{enumerate}

	\item Theoretical estimate: The numerator of $\beta$ is apparently less than the valley splitting $\tbkt{\lambda z}{\Delta V}{\lambda \bar{z}}$, where $\ket{\lambda z}$ is the ground orbital state.  Since the valley splitting gets smaller as the interface roughness increases, while the orbital energy splitting stays the same, it appears that $\beta$ will be substantially less than one for a rough interface.
	\item Experimental: In terms of results for both magnitude of valley splitting and orbital energy splitting we can point to one group which has measured the valley splitting to be between 0.3 and 0.8 meV\cite{Yang13a} and the orbital energy splitting to be as large as 8 meV\cite{Yang13a}.

\end{enumerate}

\subsection{Files}

All pathnames are relative to L:/internal/SET\_team/Neil/other peoples documents/Culcer/valleys 14\_5

\begin{description}

	\item[Text:] /manuscript/text/Culcer\_valley\_phase\_v\_2.tex.
	\item[All figures] /manuscript/figures/15\_11 portrait Culcer phase figures.pptx.
	\item[Fig 1] Main: /manuscript/figures/Delta phi plots/Step\_NonVicinal.dat
           and Step\_Vicinal.dat using do\_plot\_step\_vicinal.m.
	\item[Fig 2] Main: /manuscript/figures/Delta phi plots/Sin\_NonVicinal.dat
           and Sin\_Vicinal.dat using do\_plot\_sin\_vicinal.m.
	\item[Fig 3 both panels] manuscript/figures/voltage control/VoltageControl\_XL10.dat and
          \_XL0.dat using do\_plot\_voltage\_control.m.
	\item[Fig S1] /manuscript/figures/15\_11 portrait Culcer phase figures.pptx.
	\item[Fig S2] /manuscript/figures/smoothed for Supp/xi\_x\_steps\_tilt\_0\_rand\_1.dat,
          using do\_plot\_smoothed\_unsmoothed.m, followed by xli, cla, plot(x, y, 'k-')

\end{description}

\end{document}

%% file: abbrev.tex
\newcommand{\sub} [2] {\mbox{$#1_{\mbox{\scriptsize{#2}}}$}}

\newcommand {\bra} [1] {\langle #1 |}
\newcommand {\ket} [1] {| #1 \rangle}

\newcommand {\tbkt} [3] {\langle #1 | #2 | #3 \rangle}
\newcommand {\pd} [2] {\frac{\partial #1}{\partial #2}}

 \newcommand {\beq}{\begin{equation}}
\newcommand {\eeq}{\end{equation}}

\def\bra#1{\mbox{$\left< #1\right|$}}
\def\ket#1{\mbox{$\left| #1\right>$}}


%% file: arXiv_Culcer_valley_phase_v_5.bbl
\begin{thebibliography}{38}
\expandafter\ifx\csname natexlab\endcsname\relax\def\natexlab#1{#1}\fi
\expandafter\ifx\csname bibnamefont\endcsname\relax
  \def\bibnamefont#1{#1}\fi
\expandafter\ifx\csname bibfnamefont\endcsname\relax
  \def\bibfnamefont#1{#1}\fi
\expandafter\ifx\csname citenamefont\endcsname\relax
  \def\citenamefont#1{#1}\fi
\expandafter\ifx\csname url\endcsname\relax
  \def\url#1{\texttt{#1}}\fi
\expandafter\ifx\csname urlprefix\endcsname\relax\def\urlprefix{URL }\fi
\providecommand{\bibinfo}[2]{#2}
\providecommand{\eprint}[2][]{\url{#2}}

\bibitem[{\citenamefont{Zwanenburg et~al.}(2013)}]{Zwanenburg13a}
\bibinfo{author}{\bibfnamefont{F.~A.} \bibnamefont{Zwanenburg}}
  \bibnamefont{et~al.}, \bibinfo{journal}{Rev. Mod. Phys.}
  \textbf{\bibinfo{volume}{85}}, \bibinfo{pages}{961 } (\bibinfo{year}{2013}).

\bibitem[{\citenamefont{Morton et~al.}(2011)\citenamefont{Morton, McCamey,
  Eriksson, and Lyon}}]{Morton11a}
\bibinfo{author}{\bibfnamefont{J.~J.~L.} \bibnamefont{Morton}},
  \bibinfo{author}{\bibfnamefont{D.~R.} \bibnamefont{McCamey}},
  \bibinfo{author}{\bibfnamefont{M.~A.} \bibnamefont{Eriksson}},
  \bibnamefont{and} \bibinfo{author}{\bibfnamefont{S.~A.} \bibnamefont{Lyon}},
  \bibinfo{journal}{Nature} \textbf{\bibinfo{volume}{479}}, \bibinfo{pages}{345
  } (\bibinfo{year}{2011}).

\bibitem[{\citenamefont{Scarlino et~al.}(2015)}]{Scarlino15a}
\bibinfo{author}{\bibfnamefont{P.}~\bibnamefont{Scarlino}}
  \bibnamefont{et~al.}, \bibinfo{journal}{Phys. Rev. Lett.}
  \textbf{\bibinfo{volume}{115}}, \bibinfo{pages}{106802}
  (\bibinfo{year}{2015}).

\bibitem[{\citenamefont{Kawakami et~al.}(2014)}]{Kawakami14a}
\bibinfo{author}{\bibfnamefont{E.}~\bibnamefont{Kawakami}}
  \bibnamefont{et~al.}, \bibinfo{journal}{Nature Nanotech.}
  \textbf{\bibinfo{volume}{9}}, \bibinfo{pages}{666 } (\bibinfo{year}{2014}).

\bibitem[{\citenamefont{Veldhorst et~al.}(2014)}]{Veldhorst14a}
\bibinfo{author}{\bibfnamefont{M.}~\bibnamefont{Veldhorst}}
  \bibnamefont{et~al.}, \bibinfo{journal}{Nature Nanotech.}
  \textbf{\bibinfo{volume}{9}}, \bibinfo{pages}{981 } (\bibinfo{year}{2014}).

\bibitem[{\citenamefont{Veldhorst et~al.}(2015)}]{Veldhorst15a}
\bibinfo{author}{\bibfnamefont{M.}~\bibnamefont{Veldhorst}}
  \bibnamefont{et~al.}, \bibinfo{journal}{Nature}
  \textbf{\bibinfo{volume}{526}}, \bibinfo{pages}{410 } (\bibinfo{year}{2015}).

\bibitem[{\citenamefont{Yang et~al.}(2013)}]{Yang13a}
\bibinfo{author}{\bibfnamefont{C.~H.} \bibnamefont{Yang}} \bibnamefont{et~al.},
  \bibinfo{journal}{Nature Comm.} \textbf{\bibinfo{volume}{4}},
  \bibinfo{pages}{2069} (\bibinfo{year}{2013}).

\bibitem[{\citenamefont{Hao et~al.}(2014)}]{Hao13a}
\bibinfo{author}{\bibfnamefont{X.}~\bibnamefont{Hao}} \bibnamefont{et~al.},
  \bibinfo{journal}{Nature Comm.} \textbf{\bibinfo{volume}{5}},
  \bibinfo{pages}{3860} (\bibinfo{year}{2014}).

\bibitem[{\citenamefont{Goswami et~al.}(2007)\citenamefont{Goswami, Slinker,
  Friesen, McGuire, Truitt, Tahan, Klein, Chu, Mooney, van~der Weide
  et~al.}}]{Goswami07a}
\bibinfo{author}{\bibfnamefont{S.}~\bibnamefont{Goswami}},
  \bibinfo{author}{\bibfnamefont{K.~A.} \bibnamefont{Slinker}},
  \bibinfo{author}{\bibfnamefont{M.}~\bibnamefont{Friesen}},
  \bibinfo{author}{\bibfnamefont{L.~M.} \bibnamefont{McGuire}},
  \bibinfo{author}{\bibfnamefont{J.~L.} \bibnamefont{Truitt}},
  \bibinfo{author}{\bibfnamefont{C.}~\bibnamefont{Tahan}},
  \bibinfo{author}{\bibfnamefont{L.~J.} \bibnamefont{Klein}},
  \bibinfo{author}{\bibfnamefont{J.~O.} \bibnamefont{Chu}},
  \bibinfo{author}{\bibfnamefont{P.~M.} \bibnamefont{Mooney}},
  \bibinfo{author}{\bibfnamefont{V.~W.} \bibnamefont{van~der Weide}},
  \bibnamefont{et~al.}, \bibinfo{journal}{Nature Physics}
  \textbf{\bibinfo{volume}{3}}, \bibinfo{pages}{41 } (\bibinfo{year}{2007}).

\bibitem[{\citenamefont{Shi et~al.}(2011)}]{Shi11a}
\bibinfo{author}{\bibfnamefont{Z.}~\bibnamefont{Shi}} \bibnamefont{et~al.},
  \bibinfo{journal}{Appl. Phys. Lett.} \textbf{\bibinfo{volume}{99}},
  \bibinfo{pages}{233108} (\bibinfo{year}{2011}).

\bibitem[{\citenamefont{Boykin et~al.}(2004{\natexlab{a}})\citenamefont{Boykin,
  Klimeck, Eriksson, Friesen, Coppersmith, von Allmen, Oyafuso, and
  Lee}}]{Boykin04a}
\bibinfo{author}{\bibfnamefont{T.~B.} \bibnamefont{Boykin}},
  \bibinfo{author}{\bibfnamefont{G.}~\bibnamefont{Klimeck}},
  \bibinfo{author}{\bibfnamefont{M.~A.} \bibnamefont{Eriksson}},
  \bibinfo{author}{\bibfnamefont{M.}~\bibnamefont{Friesen}},
  \bibinfo{author}{\bibfnamefont{S.~N.} \bibnamefont{Coppersmith}},
  \bibinfo{author}{\bibfnamefont{P.}~\bibnamefont{von Allmen}},
  \bibinfo{author}{\bibfnamefont{F.}~\bibnamefont{Oyafuso}}, \bibnamefont{and}
  \bibinfo{author}{\bibfnamefont{S.}~\bibnamefont{Lee}},
  \bibinfo{journal}{Applied Physics Letters} \textbf{\bibinfo{volume}{84}},
  \bibinfo{pages}{115} (\bibinfo{year}{2004}{\natexlab{a}}),
  \urlprefix\url{http://scitation.aip.org/content/aip/journal/apl/84/1/10.1063/1.1637718;jsessionid=eQPbre3FvIyOPAdejAVTNIWa.x-aip-live-03}.

\bibitem[{\citenamefont{Boykin et~al.}(2004{\natexlab{b}})\citenamefont{Boykin,
  Klimeck, Friesen, Coppersmith, von Allmen, Oyafuso, and Lee}}]{Boykin04b}
\bibinfo{author}{\bibfnamefont{T.~B.} \bibnamefont{Boykin}},
  \bibinfo{author}{\bibfnamefont{G.}~\bibnamefont{Klimeck}},
  \bibinfo{author}{\bibfnamefont{M.}~\bibnamefont{Friesen}},
  \bibinfo{author}{\bibfnamefont{S.~N.} \bibnamefont{Coppersmith}},
  \bibinfo{author}{\bibfnamefont{P.}~\bibnamefont{von Allmen}},
  \bibinfo{author}{\bibfnamefont{F.}~\bibnamefont{Oyafuso}}, \bibnamefont{and}
  \bibinfo{author}{\bibfnamefont{S.}~\bibnamefont{Lee}},
  \bibinfo{journal}{Phys. Rev. B} \textbf{\bibinfo{volume}{70}},
  \bibinfo{pages}{165325} (\bibinfo{year}{2004}{\natexlab{b}}),
  \urlprefix\url{http://link.aps.org/doi/10.1103/PhysRevB.70.165325}.

\bibitem[{\citenamefont{Boykin et~al.}(2008)\citenamefont{Boykin, Kharche, and
  Klimeck}}]{Boykin06a}
\bibinfo{author}{\bibfnamefont{T.~B.} \bibnamefont{Boykin}},
  \bibinfo{author}{\bibfnamefont{N.}~\bibnamefont{Kharche}}, \bibnamefont{and}
  \bibinfo{author}{\bibfnamefont{G.}~\bibnamefont{Klimeck}},
  \bibinfo{journal}{Phys. Rev. B} \textbf{\bibinfo{volume}{77}},
  \bibinfo{pages}{245320} (\bibinfo{year}{2008}),
  \urlprefix\url{http://link.aps.org/doi/10.1103/PhysRevB.77.245320}.

\bibitem[{\citenamefont{Friesen and Coppersmith}(2010)}]{Friesen10a}
\bibinfo{author}{\bibfnamefont{M.}~\bibnamefont{Friesen}} \bibnamefont{and}
  \bibinfo{author}{\bibfnamefont{S.~N.} \bibnamefont{Coppersmith}},
  \bibinfo{journal}{Phys. Rev. B} \textbf{\bibinfo{volume}{81}},
  \bibinfo{pages}{115324} (\bibinfo{year}{2010}),
  \urlprefix\url{http://link.aps.org/doi/10.1103/PhysRevB.81.115324}.

\bibitem[{\citenamefont{Saraiva et~al.}(2009)\citenamefont{Saraiva, Calder\'on,
  Hu, Das~Sarma, and Koiller}}]{Saraiva09a}
\bibinfo{author}{\bibfnamefont{A.~L.} \bibnamefont{Saraiva}},
  \bibinfo{author}{\bibfnamefont{M.~J.} \bibnamefont{Calder\'on}},
  \bibinfo{author}{\bibfnamefont{X.}~\bibnamefont{Hu}},
  \bibinfo{author}{\bibfnamefont{S.}~\bibnamefont{Das~Sarma}},
  \bibnamefont{and} \bibinfo{author}{\bibfnamefont{B.}~\bibnamefont{Koiller}},
  \bibinfo{journal}{Phys. Rev. B} \textbf{\bibinfo{volume}{80}},
  \bibinfo{pages}{081305} (\bibinfo{year}{2009}),
  \urlprefix\url{http://link.aps.org/doi/10.1103/PhysRevB.80.081305}.

\bibitem[{\citenamefont{Saraiva et~al.}(2011)\citenamefont{Saraiva, Calder\'on,
  Capaz, Hu, Das~Sarma, and Koiller}}]{Saraiva11a}
\bibinfo{author}{\bibfnamefont{A.~L.} \bibnamefont{Saraiva}},
  \bibinfo{author}{\bibfnamefont{M.~J.} \bibnamefont{Calder\'on}},
  \bibinfo{author}{\bibfnamefont{R.~B.} \bibnamefont{Capaz}},
  \bibinfo{author}{\bibfnamefont{X.}~\bibnamefont{Hu}},
  \bibinfo{author}{\bibfnamefont{S.}~\bibnamefont{Das~Sarma}},
  \bibnamefont{and} \bibinfo{author}{\bibfnamefont{B.}~\bibnamefont{Koiller}},
  \bibinfo{journal}{Phys. Rev. B} \textbf{\bibinfo{volume}{84}},
  \bibinfo{pages}{155320} (\bibinfo{year}{2011}),
  \urlprefix\url{http://link.aps.org/doi/10.1103/PhysRevB.84.155320}.

\bibitem[{\citenamefont{Srinivasan et~al.}(2008)\citenamefont{Srinivasan,
  Klimeck, and Rokhinson}}]{Srinivasan08a}
\bibinfo{author}{\bibfnamefont{S.}~\bibnamefont{Srinivasan}},
  \bibinfo{author}{\bibfnamefont{G.}~\bibnamefont{Klimeck}}, \bibnamefont{and}
  \bibinfo{author}{\bibfnamefont{L.~P.} \bibnamefont{Rokhinson}},
  \bibinfo{journal}{Applied Physics Letters} \textbf{\bibinfo{volume}{93}},
  \bibinfo{pages}{112102} (\bibinfo{year}{2008}),
  \urlprefix\url{http://scitation.aip.org/content/aip/journal/apl/93/11/10.1063/1.2981577;jsessionid=eQPbre3FvIyOPAdejAVTNIWa.x-aip-live-03}.

\bibitem[{\citenamefont{Gamble et~al.}(2013)\citenamefont{Gamble, Eriksson,
  Coppersmith, and Friesen}}]{Gamble13a}
\bibinfo{author}{\bibfnamefont{J.~K.} \bibnamefont{Gamble}},
  \bibinfo{author}{\bibfnamefont{M.~A.} \bibnamefont{Eriksson}},
  \bibinfo{author}{\bibfnamefont{S.~N.} \bibnamefont{Coppersmith}},
  \bibnamefont{and} \bibinfo{author}{\bibfnamefont{M.}~\bibnamefont{Friesen}},
  \bibinfo{journal}{Phys. Rev. B} \textbf{\bibinfo{volume}{88}},
  \bibinfo{pages}{035310} (\bibinfo{year}{2013}),
  \urlprefix\url{http://link.aps.org/doi/10.1103/PhysRevB.88.035310}.

\bibitem[{\citenamefont{Culcer et~al.}(2010{\natexlab{a}})}]{Culcer10b}
\bibinfo{author}{\bibfnamefont{D.}~\bibnamefont{Culcer}} \bibnamefont{et~al.},
  \bibinfo{journal}{Phys. Rev. B} \textbf{\bibinfo{volume}{82}},
  \bibinfo{pages}{205315} (\bibinfo{year}{2010}{\natexlab{a}}).

\bibitem[{\citenamefont{Friesen et~al.}(2007)\citenamefont{Friesen, Chutia,
  Tahan, and Coppersmith}}]{Friesen07a}
\bibinfo{author}{\bibfnamefont{M.}~\bibnamefont{Friesen}},
  \bibinfo{author}{\bibfnamefont{S.}~\bibnamefont{Chutia}},
  \bibinfo{author}{\bibfnamefont{C.}~\bibnamefont{Tahan}}, \bibnamefont{and}
  \bibinfo{author}{\bibfnamefont{S.~N.} \bibnamefont{Coppersmith}},
  \bibinfo{journal}{Phys. Rev. B} \textbf{\bibinfo{volume}{75}},
  \bibinfo{pages}{115318} (\bibinfo{year}{2007}),
  \urlprefix\url{http://link.aps.org/doi/10.1103/PhysRevB.75.115318}.

\bibitem[{\citenamefont{Friesen et~al.}(2006)\citenamefont{Friesen, Eriksson,
  and Coppersmith}}]{Friesen06a}
\bibinfo{author}{\bibfnamefont{M.}~\bibnamefont{Friesen}},
  \bibinfo{author}{\bibfnamefont{M.~A.} \bibnamefont{Eriksson}},
  \bibnamefont{and} \bibinfo{author}{\bibfnamefont{S.~N.}
  \bibnamefont{Coppersmith}}, \bibinfo{journal}{Applied Physics Letters}
  \textbf{\bibinfo{volume}{89}}, \bibinfo{pages}{202106}
  (\bibinfo{year}{2006}),
  \urlprefix\url{http://scitation.aip.org/content/aip/journal/apl/89/20/10.1063/1.2387975;jsessionid=e8uHB9fAyj0HD4RgKZM7PzA+.x-aip-live-03}.

\bibitem[{\citenamefont{Wu and Culcer}(2012)}]{Wu12a}
\bibinfo{author}{\bibfnamefont{Y.}~\bibnamefont{Wu}} \bibnamefont{and}
  \bibinfo{author}{\bibfnamefont{D.}~\bibnamefont{Culcer}},
  \bibinfo{journal}{Phys. Rev. B} \textbf{\bibinfo{volume}{86}},
  \bibinfo{pages}{035321} (\bibinfo{year}{2012}),
  \urlprefix\url{http://link.aps.org/doi/10.1103/PhysRevB.86.035321}.

\bibitem[{\citenamefont{Culcer et~al.}(2010{\natexlab{b}})}]{Culcer10a}
\bibinfo{author}{\bibfnamefont{D.}~\bibnamefont{Culcer}} \bibnamefont{et~al.},
  \bibinfo{journal}{Phys. Rev. B} \textbf{\bibinfo{volume}{82}},
  \bibinfo{pages}{155312} (\bibinfo{year}{2010}{\natexlab{b}}).

\bibitem[{\citenamefont{Huang et~al.}(2017)\citenamefont{Huang, Veldhorst,
  Zimmerman, Dzurak, and Culcer}}]{Huang16a}
\bibinfo{author}{\bibfnamefont{W.}~\bibnamefont{Huang}},
  \bibinfo{author}{\bibfnamefont{M.}~\bibnamefont{Veldhorst}},
  \bibinfo{author}{\bibfnamefont{N.~M.} \bibnamefont{Zimmerman}},
  \bibinfo{author}{\bibfnamefont{A.~S.} \bibnamefont{Dzurak}},
  \bibnamefont{and} \bibinfo{author}{\bibfnamefont{D.}~\bibnamefont{Culcer}},
  \bibinfo{journal}{Phys. Rev. B} \textbf{\bibinfo{volume}{95}},
  \bibinfo{pages}{075403} (\bibinfo{year}{2017}).

\bibitem[{\citenamefont{Boross et~al.}(2016)\citenamefont{Boross, Széchenyi,
  Culcer, and Pályi}}]{Boross16a}
\bibinfo{author}{\bibfnamefont{P.}~\bibnamefont{Boross}},
  \bibinfo{author}{\bibfnamefont{G.}~\bibnamefont{Széchenyi}},
  \bibinfo{author}{\bibfnamefont{D.}~\bibnamefont{Culcer}}, \bibnamefont{and}
  \bibinfo{author}{\bibfnamefont{A.}~\bibnamefont{Pályi}},
  \bibinfo{journal}{Phys. Rev. B} \textbf{\bibinfo{volume}{94}},
  \bibinfo{pages}{035438} (\bibinfo{year}{2016}).

\bibitem[{\citenamefont{yuan Shiau et~al.}(2007)\citenamefont{yuan Shiau,
  Chutia, and Joynt}}]{Shiau07a}
\bibinfo{author}{\bibfnamefont{S.}~\bibnamefont{yuan Shiau}},
  \bibinfo{author}{\bibfnamefont{S.}~\bibnamefont{Chutia}}, \bibnamefont{and}
  \bibinfo{author}{\bibfnamefont{R.}~\bibnamefont{Joynt}},
  \bibinfo{journal}{Phys. Rev. B} \textbf{\bibinfo{volume}{75}},
  \bibinfo{pages}{195345} (\bibinfo{year}{2007}).

\bibitem[{\citenamefont{Rahman et~al.}(2012)\citenamefont{Rahman, Nielsen,
  Muller, and Carroll}}]{Rahman2012a}
\bibinfo{author}{\bibfnamefont{R.}~\bibnamefont{Rahman}},
  \bibinfo{author}{\bibfnamefont{E.}~\bibnamefont{Nielsen}},
  \bibinfo{author}{\bibfnamefont{R.~P.} \bibnamefont{Muller}},
  \bibnamefont{and} \bibinfo{author}{\bibfnamefont{M.~S.}
  \bibnamefont{Carroll}}, \bibinfo{journal}{Phys. Rev. B}
  \textbf{\bibinfo{volume}{85}}, \bibinfo{pages}{125423}
  (\bibinfo{year}{2012}), \bibinfo{note}{voltage controlled exchange energies
  of a two-electron silicon double quantum dot with and without charge defects
  in the dielectric}.

\bibitem[{\citenamefont{Ikarashi and Watanabe}(2000)}]{ikarashi00a}
\bibinfo{author}{\bibfnamefont{N.}~\bibnamefont{Ikarashi}} \bibnamefont{and}
  \bibinfo{author}{\bibfnamefont{K.}~\bibnamefont{Watanabe}},
  \bibinfo{journal}{Jpn. J. Appl. Phys.} \textbf{\bibinfo{volume}{39}},
  \bibinfo{pages}{1278 } (\bibinfo{year}{2000}).

\bibitem[{\citenamefont{Krivanek and Mazur}(1980)}]{krivanek80a}
\bibinfo{author}{\bibfnamefont{O.~L.} \bibnamefont{Krivanek}} \bibnamefont{and}
  \bibinfo{author}{\bibfnamefont{J.~H.} \bibnamefont{Mazur}},
  \bibinfo{journal}{Appl. Phys. Lett.} \textbf{\bibinfo{volume}{37}},
  \bibinfo{pages}{392 } (\bibinfo{year}{1980}).

\bibitem[{\citenamefont{Anderson et~al.}(1993)}]{anderson93a}
\bibinfo{author}{\bibfnamefont{W.~R.} \bibnamefont{Anderson}}
  \bibnamefont{et~al.}, \bibinfo{journal}{Micro. Eng.}
  \textbf{\bibinfo{volume}{22}}, \bibinfo{pages}{43 } (\bibinfo{year}{1993}).

\bibitem[{\citenamefont{Weber et~al.}(2014)}]{Weber14a}
\bibinfo{author}{\bibfnamefont{B.}~\bibnamefont{Weber}} \bibnamefont{et~al.},
  \bibinfo{journal}{Nature Nano.} \textbf{\bibinfo{volume}{9}},
  \bibinfo{pages}{430} (\bibinfo{year}{2014}).

\bibitem[{\citenamefont{Petta et~al.}(2005)}]{Petta05a}
\bibinfo{author}{\bibfnamefont{J.~R.} \bibnamefont{Petta}}
  \bibnamefont{et~al.}, \bibinfo{journal}{Science}
  \textbf{\bibinfo{volume}{309}}, \bibinfo{pages}{2180 }
  (\bibinfo{year}{2005}).

\bibitem[{\citenamefont{Lai et~al.}(2011)}]{Lai11a}
\bibinfo{author}{\bibfnamefont{N.~S.} \bibnamefont{Lai}} \bibnamefont{et~al.},
  \bibinfo{journal}{Sci. Rep.} \textbf{\bibinfo{volume}{1}},
  \bibinfo{pages}{110} (\bibinfo{year}{2011}).

\bibitem[{\citenamefont{Shaji et~al.}(2008)}]{Shaji08a}
\bibinfo{author}{\bibfnamefont{N.}~\bibnamefont{Shaji}} \bibnamefont{et~al.},
  \bibinfo{journal}{Nature Physics} \textbf{\bibinfo{volume}{4}},
  \bibinfo{pages}{540 } (\bibinfo{year}{2008}).

\bibitem[{\citenamefont{Perron et~al.}(2016)}]{Perron16a}
\bibinfo{author}{\bibfnamefont{J.~K.} \bibnamefont{Perron}}
  \bibnamefont{et~al.} (\bibinfo{year}{2016}), \bibinfo{note}{unpublished}.

\bibitem[{\citenamefont{Li et~al.}(2011)}]{Li11a}
\bibinfo{author}{\bibfnamefont{K.}~\bibnamefont{Li}} \bibnamefont{et~al.},
  \bibinfo{journal}{J. Vac. Sci. Technol. B} \textbf{\bibinfo{volume}{29}},
  \bibinfo{pages}{041806} (\bibinfo{year}{2011}).

\bibitem[{\citenamefont{Tanaka et~al.}(1996)}]{Tanaka96a}
\bibinfo{author}{\bibfnamefont{S.}~\bibnamefont{Tanaka}} \bibnamefont{et~al.},
  \bibinfo{journal}{Appl. Phys. Lett.} \textbf{\bibinfo{volume}{69}},
  \bibinfo{pages}{1235 } (\bibinfo{year}{1996}).

\bibitem[{\citenamefont{Streetman}(1995)}]{Streetman95a}
\bibinfo{author}{\bibfnamefont{B.~G.} \bibnamefont{Streetman}},
  \emph{\bibinfo{title}{{Solid State Electronic Devices}}}
  (\bibinfo{publisher}{Prentice Hall}, \bibinfo{address}{Englewood Cliffs, NJ,
  USA}, \bibinfo{year}{1995}).

\end{thebibliography}
